\documentclass[twocolumn,showpacs,pre,floatfix,amsmath,amssymb,notitlepage,superscriptaddress]{revtex4-1}
\usepackage{epsfig}
\usepackage{subfigure}
\usepackage{amsmath}
\usepackage{color}
\usepackage{amssymb}
\usepackage{setspace}
\usepackage{graphicx}
\usepackage{dcolumn}
\usepackage{bm}
\usepackage{times}
\usepackage{enumerate}
\usepackage{footnote}
\input epsf

\usepackage{algorithm}	
\usepackage{algpseudocode}

\newcommand{\appropto}{\mathrel{\vcenter{
  \offinterlineskip\halign{\hfil$##$\cr
    \propto\cr\noalign{\kern2pt}\sim\cr\noalign{\kern-2pt}}}}}

\graphicspath{{figures/}}

\begin{document}

\author{Juan G. Restrepo}
\affiliation{Department of Applied Mathematics, University of Colorado at Boulder, Boulder, CO 80309, USA}

\author{Clayton P. Byers}
\affiliation{Department of Engineering, Trinity College, Hartford, CT, 06106, USA}

\author{Per Sebastian Skardal}
\affiliation{Department of Mathematics, Trinity College, Hartford, CT, 06106, USA}

\title{Suppressing unknown disturbances to dynamical systems using machine learning}
\date{\today}

\begin{abstract}
Identifying and suppressing unknown disturbances to dynamical systems is a problem with applications in many different fields. Here we present a model-free method to identify and suppress an unknown disturbance to an unknown system based only on previous observations of the system under the influence of a known forcing function. We find that, under very mild restrictions on the training function, our method is able to robustly identify and suppress a large class of unknown disturbances. We illustrate our scheme with the identification of both deterministic and stochastic unknown disturbances to an analog electric chaotic circuit and with numerical examples where a chaotic disturbance to various chaotic dynamical systems is identified and suppressed.
\end{abstract}


\maketitle

\section*{Introduction}\label{sec:01}

Identifying and suppressing an unknown disturbance to a dynamical system is a problem with many existing and potential applications in engineering \cite{nudell2013graph, upadhyaya2015power, mathew2016pmu, chen2015disturbance, ferreira2016survey, lee2018data, wang2019detection, delabays2021locating,delabays2022locating}, ecology \cite{meurant2012ecology,battisti2016introduction}, fluid mechanics \cite{bewley1998optimal, bewley2000general,juillet2014experimental}, and climate change \cite{verbesselt2010phenological}. Traditional control theory disturbance identification and suppression techniques usually assume either an existing model for the dynamical system, linearity, or that the disturbance can be observed (for reviews of existing methods see, for example, Refs.~\cite{canfield2003active, chen2015disturbance}). In this Article we present a method for real-time disturbance identification and suppression that relies solely on observations of the dynamical system when forced with a known training forcing function. Our method is based on the application of machine-learning techniques to dynamical systems. Such techniques have found many applications, including the forecast of chaotic spatiotemporal \cite{pathak2018model} and networked \cite{srinivasan2022parallel} dynamics, estimation of dynamical invariants from data \cite{pathak2017using}, control of chaos \cite{canaday2021model}, network structure inference \cite{banerjee2019using}, and prediction of extreme events \cite{pyragas2020using} and crises in non-stationary dynamical systems \cite{kong2023reservoir,patel2023using}. For a review of other applications and techniques, see Refs.~\cite{tanaka2019recent,nakajima2021reservoir,carrol2019Chaos}. In most of these previous works, a machine learning framework is trained to replicate the nonlinear dynamics of the system based on a sufficiently long time series of the dynamics. 

In this Article we use machine learning to identify and subsequently suppress an unknown disturbance. Without knowledge of an underlying model for the dynamical system, and only based on observations of the system under a suitable known forcing function, our method allows us to reliably identify and suppress a large class of disturbances. Recent work \cite{mandal2023learning} considers the problem of predicting the response of a system based on knowledge of the forcing (the disturbance) and the system's response after training with known functions. That problem can be thought of as the ``forward'' problem, while the problem addressed here can be considered as the ``inverse'' problem. While both approaches are complementary, they apply to very different situations. In addition to this fundamental difference, the main additional differences between our results and those of Ref.~\cite{mandal2023learning} are that we present a method to suppress the unknown forcing, that our method works for stochastic signals, and that we show that the training functions can be extremely simple (e.g., piecewise constant functions). We also demonstrate our method with an experimental analog chaotic circuit in addition to numerical simulations.

\section*{Results}\label{sec:02}

\subsection*{System Setup, Disturbances, and Reservoir Computers}\label{sec:02:01}

Consider an $N$-dimensional dynamical system
\begin{align}
\frac{d{\bf x}}{dt} = {\bf F}({\bf x}) +{\bf g}(t),\label{eq}
\end{align}
where ${\bf x} \in \mathbb{R}^N$ is the state vector, ${\bf F} \in \mathbb{R}^N$ represents the intrinsic dynamics of the system, and ${\bf g}(t)\in \mathbb{R}^N$ represents an unknown (and usually undesired) disturbance. Our goal is to develop a scheme by which the disturbance can be identified and the system can be brought approximately to satisfy the undisturbed dynamics $d{\bf x}/dt = {\bf F}({\bf x})$. We assume that we can observe  the state vector ${\bf x}$, but  we don't need to assume knowledge of the intrinsic dynamics ${\bf F}$ or the disturbance function ${\bf g}$. Assuming that we can force the system with a  known {\it training forcing function} ${\bf f}(t)$, as
\begin{align}
\frac{d\hat{ \bf x}}{dt} = {\bf F}(\hat {\bf x}) +{\bf f}(t),\label{train}
\end{align}
and observe $\hat {\bf x}(t)$ for a long enough time, our goal is to train a machine learning system to approximate ${\bf f}(t)$ given $\hat  {\bf x}(t)$, and subsequently infer ${\bf g}(t)$ from observations of ${\bf x}(t)$ obtained from system (\ref{eq}). As we will show below, we find that we can recover a large class of forcing functions ${\bf g}(t)$ with very mild restrictions on the choice of training functions ${\bf f}(t)$. Once we infer ${\bf g}(t)$, we implement a self-consistent control scheme to suppress it from the dynamics. Our method works when the intrinsic dynamics are chaotic, periodic, or stationary.

\begin{figure}[t]
    \centering
    \includegraphics[width = \linewidth]{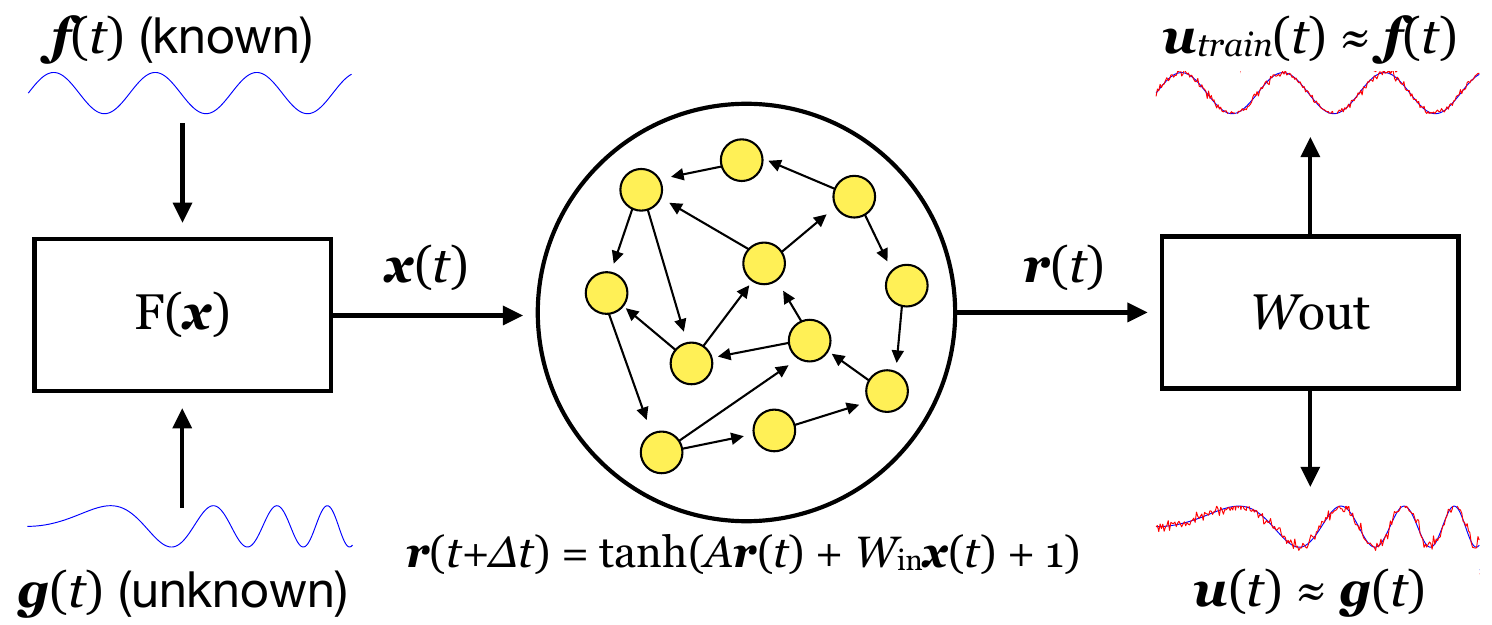}
    \caption{Schematic illustration of our method. In the training phase (top row), a nonlinear system  is forced with a training function {\bf f}(t). Observations of the forced system are used to train a reservoir  to approximate the training function. The reservoir subsequently identifies unknown disturbance functions (bottom row).}
    \label{fig1}
\end{figure}

We begin by outlining our method for identifying the unknown disturbance ${\bf g}(t)$. We will illustrate our technique using reservoir computing, a type of machine learning framework particularly suited for dynamical systems problems~\cite{nakajima2021reservoir}. In our implementation, we assume that we run the system in Eq.~(\ref{train}) during the ``training'' interval $[-T,0]$, and collect a time-series of the observed state vector $\{ \hat {\bf x}(-T),\hat {\bf x}(-T +\Delta t),\dots,\hat {\bf x}(0)\}$. These variables are fed to the reservoir, a high-dimensional dynamical system with internal variables ${\bf r}\in \mathbb{R}^M$, where $M$ is the size of the reservoir. Here, following \cite{pathak2017using,pathak2018model}, we implement the reservoir as the map
\begin{align}
{\bf r}(t+\Delta t) = \tanh[A {\bf r}(t) +W_{\text{in}} \hat{\bf x}(t) +\beta],\label{eq:03}
\end{align}
where the $M\times M$ matrix $A$ is a sparse matrix representing the internal structure of the reservoir network and the $M\times N$ matrix $W_{\text{in}}$ is a fixed input matrix. Here we choose the bias parameter $\beta = 1$, which for our purposes nearly optimizes results (see Methods). The reservoir output $\bf u$ is constructed from the internal states as
$\bf u = W_{\text{out}} {\bf r}$, where the $N\times M$  output matrix $W_{\text{out}}$ is chosen so that $\bf u$ approximates as best as possible the known training forcing function ${\bf f}(t)$. The optimization can be done by minimizing the cost function
\begin{align}
\sum_{n=0}^{T/\Delta t} \| {\bf f}(-n\Delta t) - {\bf u}(-n\Delta t) \|^2 + \lambda \text{Tr}({W_{\text{out}} W_{\text{out}}^\text{T}})
\end{align}
via a ridge regression procedure, where the constant $\lambda \geq 0$ prevents over-fitting. With this procedure, the reservoir is trained to identify the forcing function ${\bf f}(t)$ given the observed values of $\hat {\bf x}(t)$. The reservoir can then be presented with a time series of the observed variables taken from (\ref{eq}), i.e., it can be evolved as
\begin{align}
{\bf r}(t+\Delta t) = \tanh[A {\bf r}(t) +W_{\text{in}} {\bf x}(t)+1],
\end{align}
The reservoir output ${\bf u}(t) = W_{\text{out}} {\bf r}$ will be, if the method is successful, a good approximation to the unknown disturbance, ${\bf u} \approx {\bf g}$. As we will see, the reservoir robustly identifies disturbances it has not observed previously. 

\begin{figure*}[t]
    \centering
    \includegraphics[width = 0.75\linewidth]{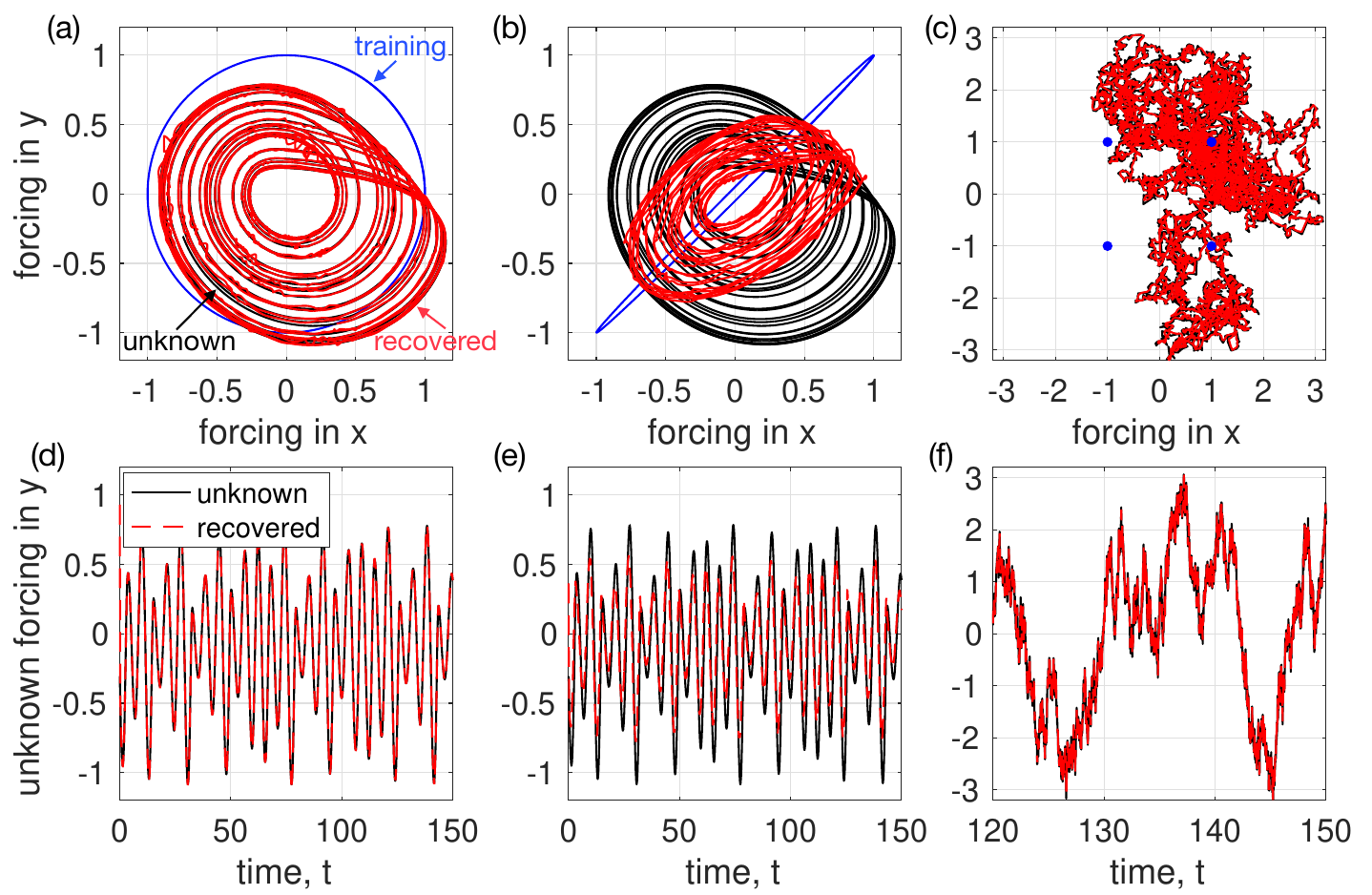}
    \caption{Identifying unknown disturbances. (Top) Unknown and reconstructed disturbance functions $[g_x(t),g_y(t)]$ (black curve) and $[u_x(t),u_y(t)]$ (red curve) along with the training forcing functions $[f_x(t),f_y(t)]$ (thick blue curves and symbols). For each case the reservoir was trained with (a) $[f_x(t), f_y(t)] = [\cos(0.05t), \sin(0.05t)]$, (b) $[f_x(t), f_y(t)] = [\cos(0.05t), \cos(0.05t-0.05)]$, and (c) $[f_x(t), f_y(t)] = [\text{sign}(\cos(0.05t)), \text{sign}(\sin(0.05t))]$. (Bottom) Time series for the unknown (solid black) and recovered (dashed red) disturbance functions in the $y$ component, $g_y(t)$ and $u_y(t)$.}
    \label{fig2}
\end{figure*}

In our numerical examples, the reservoir matrix $A$ is a random matrix of size $M = 1000$ where each entry is uniformly distributed in $[-0.5,0.5]$ with probability $6/M$ and $0$ otherwise, and rescaled so that its spectral radius is $1.2$. The input matrix $W_{\text{in}}$ is a random matrix where each entry is uniformly distributed in $[-0.01,0.01]$. The ridge regression regularization constant is $\lambda = 10^{-6}$. We train the reservoir for $T = 150$ time units and use Euler's method to solve the differential equations with a time step $\Delta t = 0.002$.

\subsection*{Simulated Examples: Deterministic and Stochastic Disturbances}\label{sec:02:02}

We first demonstrate our method with numerical simulations. For the numerical simulations, we consider a system where the intrinsic dynamics are given by the Lorenz system \cite{lorenz1963deterministic}, i.e., system (\ref{eq}) is
\begin{align}
\frac{dx_L}{dt} &= \sigma( y_L- x_L) + g_x(t),\label{l1}\\
\frac{dy_L}{dt} &=  x_L(\rho -  z_L) -y + g_y(t),\label{l2}\\
\frac{dz_L}{dt} &=  x_L  y_L - \beta  z_L + g_z(t),\label{l3}
\end{align}
with $\rho = 28$, $\sigma = 10$, and $\beta = 8/3$. For the unknown disturbance we consider two examples: (i) a deterministic forcing $[g_x,g_y,g_z]^T = [x_R/10,y_R/10,0]^T$, where $x_R(t)$ and $y_R(t)$ are the $x$ and $y$ coordinates of an auxiliary R\"ossler system \cite{rossler1976equation} (assumed to be unknown),
\begin{align}
\frac{dx_R}{dt}&= -y_R - z_R,\label{eq:r1}\\
\frac{dy_R}{dt}&=  x_R + a y_R,\label{eq:r2}\\
\frac{dz_R}{dt}&=  b + z_R(x_R-c),\label{eq:r3}
\end{align}
with $a = 0.2$, $b = 0.2$, and $c = 5.7$, and (ii) a stochastic forcing  $[x_S,y_S, 0]^T$ where both $x_S$ and $y_S$ satisfy the Langevin equations 
\begin{align}
\frac{dx_S}{dt}&=-\frac{x_S}{2}+\eta_x(t),\label{eq:s1}\\
\frac{dy_S}{dt}&=-\frac{y_S}{2}+\eta_y(t),\label{eq:s2}
\end{align}
where $\eta_x$ and $\eta_y$ are both white noise terms satisfying $\langle \eta(t)\eta(t') \rangle = 2D \delta(t-t')$, with $D = 1.25$. 

We present our results in Fig.~\ref{fig2} demonstrating the performance of the reservoir in recovering the unknown disturbance for different choices of training forcing function $[f_x(t) f_y(t), 0]^T$. (For simplicity of visualization we assume it is known that the forcing in the $z$ coordinate is zero). Along the top row, i.e., panels (a)--(c), we plot the trajectory of the unknown disturbance functions to be reconstructed as black curves, the reconstructed disturbances as red curves, and the training forcing functions as blue curves and circles. (Fig.~\ref{fig2}(c) only shows the last $1/8$ portion of the time-series.) From left to right we have trained the reservoirs with forcing functions consisting of a sine/cosine pair $[f_x(t), f_y(t)]^T = [\cos(t/20), \sin(t/20)]^T$, a slightly offset pair of cosine functions $[f_x(t), f_y(t)]^T = [\cos(t/20), \cos((t-1)/20)]^T$, and piecewise constant functions $[f_x(t), f_y(t)]^T = [\text{sign}(\cos(t/20)), \text{sign}(\sin(t/20))]^T$. Time series for the unknown and recovered disturbances, $g_y(t)$ and $u_y(t)$, are compared in the bottom row, (d)--(f), plotted in solid black and dashed red, respectively.

Remarkably, our results show that the reservoir can identify a chaotic or stochastic forcing function to the Lorenz system even when it was trained with a periodic function [Fig.~\ref{fig2}(a)], or a piecewise constant function with only {\it four} different values [Fig.~\ref{fig2}(c)]. 
Figure~\ref{fig2}(b) illustrates the limitations on the forcing functions used to train the reservoir. In this example, the forcing functions satisfy $f_x \approx f_y$. Given this limited training, the reservoir has trouble extrapolating to functions away from the manifold $f_x = f_y$, and the reconstruction of the disturbance suffers.

\begin{figure}[t]
    \centering
    \includegraphics[width = \linewidth]{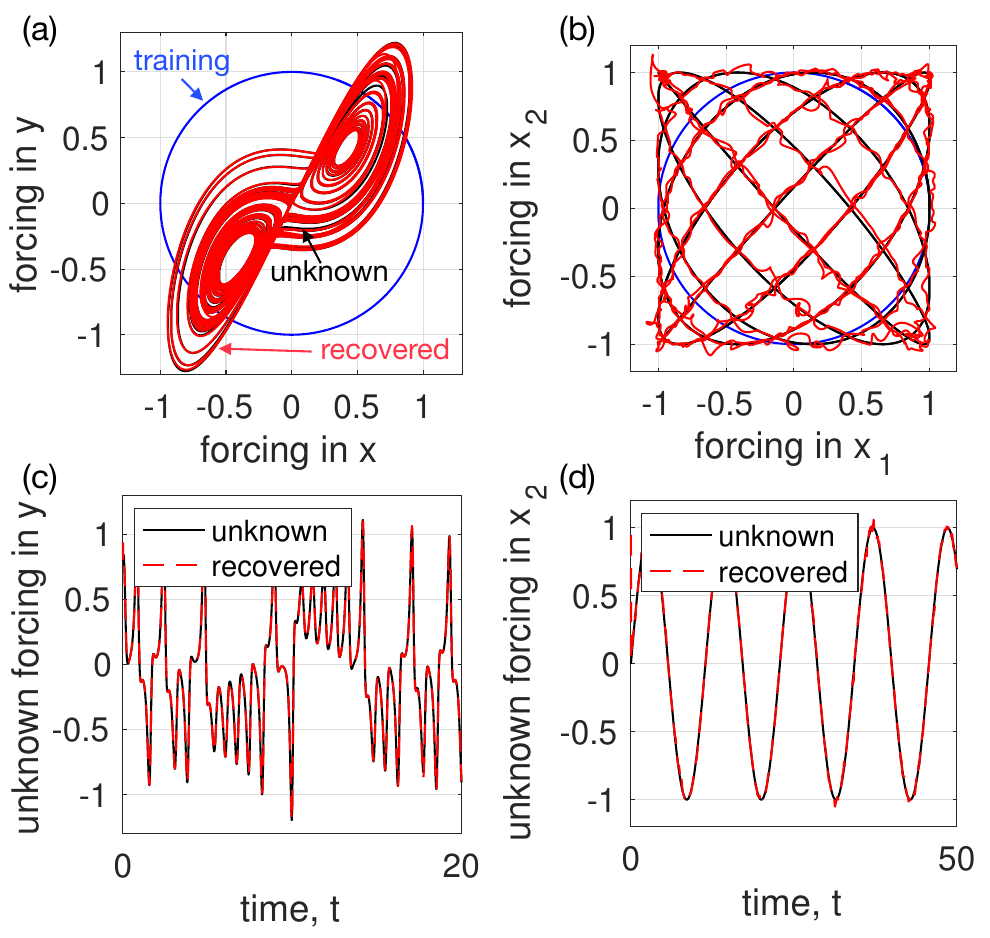}
    \caption{Identifying unknown disturbances: R\"{o}ssler and Lorenz 96 systems. (Top) Unknown and reconstructed disturbance functions $[g_x(t),g_y(t)]$ (black curve) and $[u_x(t),u_y(t)]$ (red curve) along with the training forcing functions $[f_x(t),f_y(t)]$ (blue curves). For each case the reservoir was trained with $[f_x(t), f_y(t)] = [\cos(0.05t), \sin(0.05t)]$. (Bottom) Time series for the unknown (solid black) and recovered (dashed red) disturbance functions in the second component, $g_y(t)$ and $u_y(t)$ (or $g_{x_2}(t)$ and $u_{x_2}(t)$).}
    \label{fig3}
\end{figure}

Next we present some additional results demonstrate generalizability of our mechanism. First, we consider an inverted version of our first example, namely we consider system dynamics defined by the R\"ossler system [i.e., Eqs.~(\ref{eq:r1})--(\ref{eq:r3})] that are disturbed by time series arising from the Lorenz system [i.e., Eqs.~(\ref{l1})--(\ref{l3})]. Specifically, we set $[g_x(t), g_y(t)]^T = [x_L(t)/20,y_L(t)/20]^T$ and use sinusoidal forcing, as before, for training, namely $[f_x(t), f_y(t)]^T = [\cos(t/20), \sin(t/20)]^T$. Results for this inverted example are plotted in Figs.~\ref{fig3}(a) and (c), and show good agreement between the unknown and recovered disturbances. Second, to illustrate the efficacy of the methodology in a higher-dimensional system, we consider the Lorenz 96 model~\cite{Lorenz96} whose variables $x_{i}$ for $i=1,\dots,N$ evolve according to
\begin{align}
\frac{dx_i}{dt}&= (x_{i+1} - x_{i-2})x_{i-1}-x_i + F.\label{eq:lorenz96}
\end{align}
Here we choose the dimension $N=8$ and set $F=8$ to realize high-dimensional chaos. We train the system with known forcing in the first two variables as in the prior example, $[f_{x_1}(t), f_{x_2}(t)]^T = [\cos(t/20), \sin(t/20)]^T$, then use an unknown disturbance of $[g_{x_1}(t), g_{x_2}(t)]^T = [\cos(t/2),\sin(11t/20)]^T$. Due to the high dimensionality we use a resevoir of twice the size as in other examples, namely, $M=2000$. Results for this example are plotted in Figs.~\ref{fig3}(b) and (d), and show good agreement between the unknown and recovered disturbances.

\begin{figure}[t]
    \centering
    \includegraphics[width = \linewidth]{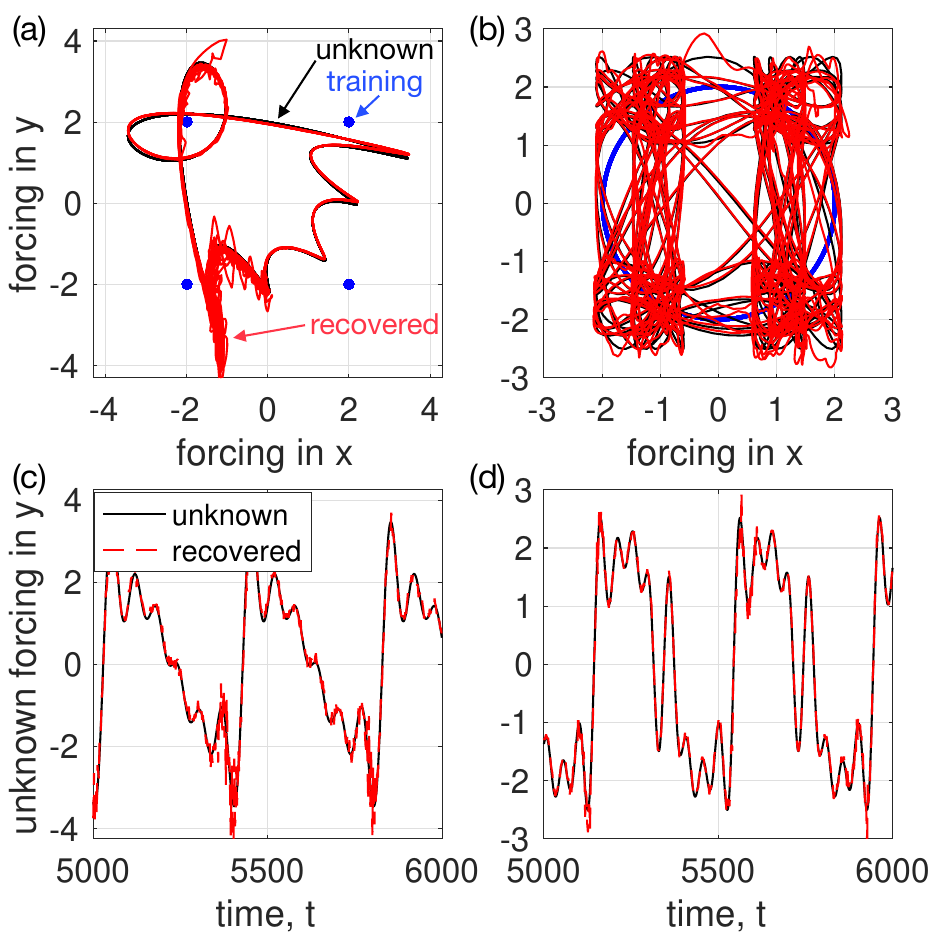}
    \caption{Experimental results. (Top) Unknown and reconstructed disturbance functions $[g_x(t),g_y(t)]$ (black curve) and $[u_x(t),u_y(t)]$ (red curve) along with the training forcing functions $[f_x(t),f_y(t)]$ (thick blue curves and symbols). For each case the reservoir was trained with (a) a 5 hz square wave out of phase by $\pi/2$ and (b) $[f_x(t), f_y(t)] = [(\cos(10\pi t)),(\sin(10\pi t))]$. (Bottom) Time series for the unknown (solid black) and recovered (dashed red) disturbance functions in the $y$ component, $g_y(t)$ and $u_y(t)$.}
    \label{fig4}
\end{figure}

\subsection*{Experimental Examples: A Chaotic Circuit}\label{sec:02:03}

In addition to the numerical simulations presented above, we also demonstrate that our method can recover unknown disturbances in an experimental setting. An analog electric circuit which reproduces the dynamics of the Lorenz equations was built following Ref.~\cite{horowitz2020art} (see Methods) and arbitrary waveform generators were used to introduce various types of additive forcing terms in both the $x$ and $y$ variables as in Eqs.~(\ref{l1})--(\ref{l2}). The circuit variables $x$, $y$, and $z$ were sampled at a rate of $10$ kHz for $20$ seconds when forced with various choices of $\bm{f}$ and $\bm{g}$. In Fig.~\ref{fig4}(a) and (b) we present the results obtained from training the reservoir using the dynamics of the circuit under piecewise constant and sinusoidal forcing, respectively, and recovering the more complicated unknown disturbance. To alleviate noise effects, the recovered disturbance is a moving average of the reservoir prediction with a window of $20$ms. Time series for the unknown disturbance and the noise-filtered recovered disturbance are shown in Figs.~\ref{fig2}(c) and (d). Despite some noise, the reservoir robustly recovers the disturbances.

An important question is what training forcing function ${\bf f}$ should one use in order to recover an {\it a priori} unknown disturbance ${\bf g}$. In our numerical experiments, we have found that the reservoir computer is able to identify disturbances with range in a region approximately $5$ times larger than the convex hull of the set $\{{\bf f}(-n \Delta t)\}_{n=0}^{T/\Delta t}$, with the same center.  This condition is very mild and can be met with a variety of simple forcing functions, for example a piecewise constant function with only {\it three values}. 
Intuitively, if the range of the training forcing function ${\bf f}$ does not contain enough information for the reservoir computer to extrapolate and infer the disturbance, the process will fail. In this Article we have not attempted a rigorous or more general analysis of the conditions that training forcing functions should satisfy, and leave this for future research. In addition to the example above where the unknown disturbance is chaotic, we have also successfully identified  temporally localized, constant, periodic, and slowly varying, non-oscillatory forcing functions ${\bf g}(t)$.

\begin{figure*}[t]
    \centering
    \includegraphics[width = 0.9\linewidth]{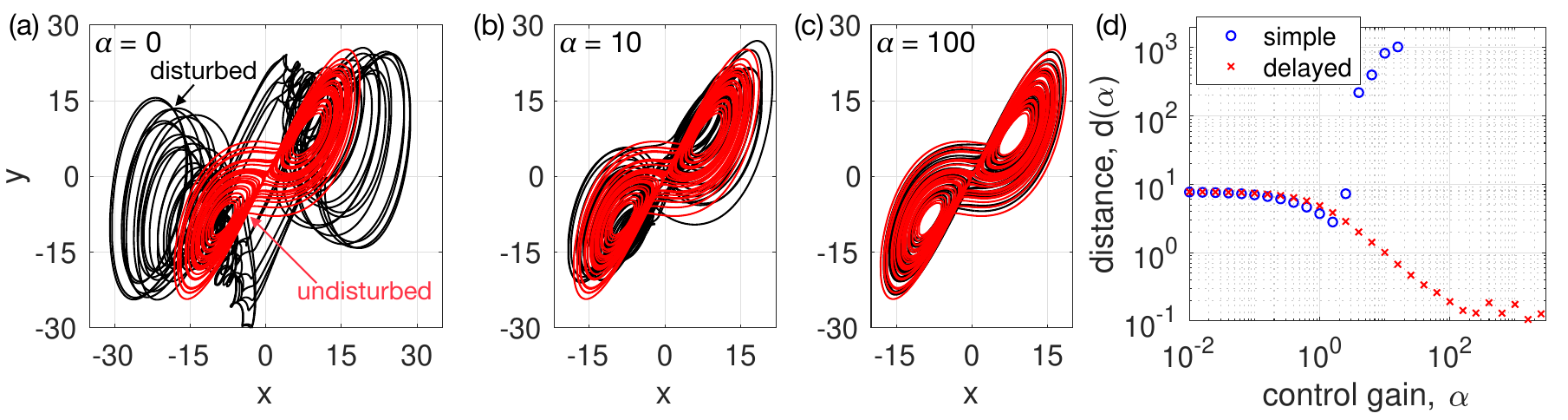}
    \caption{Suppressing unknown disturbances. (a)--(c) For control gains $\alpha=0$, $10$, and $100$, the disturbed attractor obtained from delayed control (black curves) compared to the undisturbed reference attractor (red curves). (d) As a function of the control gain $\alpha$, the average distance $d(\alpha)$ between the disturbed and original attractors as simple and delayed control (blue circles and red crosses) is applied to the disturbed attractor.}
    \label{fig5}
\end{figure*}

\subsection*{Suppression of Disturbances}\label{sec:02:04}

Next we consider the problem of suppressing the undesired disturbance function ${\bf g}(t)$ with the aim of recovering the approximate undisturbed dynamics $d{\bf x}/dt={\bf F}({\bf x})$. For this we assume that the procedure described above has been successful, and that ${\bf u}(t) =   W_{\text{out}} {\bf r}$ approximates the forcing to the system. We motivate our subsequent method by first considering a scheme where Eq.~(\ref{eq}) is modified to 
\begin{align}
\frac{d{\bf x}}{dt} = {\bf F}({\bf x}) +{\bf g}(t) - \alpha {\bf u}(t),\label{control1}
\end{align}
where $\alpha$ is the control gain, and ${\bf u}$ is obtained by feeding ${\bf x}$ to the trained reservoir. We refer to this scheme as the {\it simple control} scheme. Since the reservoir was trained to identify the forcing, in principle we have a self-consistent relationship
\begin{align}
{\bf u} \approx {\bf g} - \alpha {\bf u},\label{sc}
\end{align}
with solution
\begin{align}
{\bf u}(t) \approx \frac{{\bf g}(t)}{1+\alpha}.\label{fp}
\end{align}
The effective forcing ${\bf g}(t) - \alpha {\bf u}(t)$ in Eq.~(\ref{control1}) reduces to ${\bf g}(t)/(1+\alpha)$. In principle, then, choosing $\alpha \gg 1$ suppresses the forcing. However, this control scheme becomes unstable for moderate values of $\alpha$. To understand this, we assume momentarily that ${\bf g}$ is constant, and study the stability of the control scheme.  On a given time step, the reservoir tries to approximate the forcing in Eq.~(\ref{control1}), which is based on the previous reservoir output. Therefore, Eq.~(\ref{sc}) needs to be treated as a dynamical system. A first approximation is
\begin{align}
{\bf u}(t + \Delta t) \approx {\bf g} - \alpha {\bf u}(t),\label{ev}
\end{align}
which assumes that the reservoir approximates its own output at the previous time. In reality, the right-hand side of Eq.~(\ref{ev}) might depend on previous history. Therefore, we regard Eq.~(\ref{ev}) as a rough approximation to guide us in constructing a useful control scheme.
Under Eq.~(\ref{ev}), the fixed point $(\ref{fp})$, and therefore the control scheme, becomes unstable for $\alpha > 1$. In our example, we find numerically that the scheme becomes unstable at $\alpha \approx 2.5$, presumably due to the fact that Eq.~(\ref{ev}) is only an approximation. Additional tests using non-constant ${\bf g}$ show the same behavior.

In order to create a more robust control scheme, we modify (\ref{control1}) to
\begin{align}
\frac{d{\bf x}}{dt} &= {\bf F}({\bf x}) +{\bf g}(t) - \alpha {\bf v}(t),\label{control2}\\
\frac{d{\bf v}}{dt} &= \frac{1}{\tau}({\bf u} - {\bf v}),\label{control3}
\end{align}
where $\tau$ is a control parameter. We refer to this scheme as the {\it delayed control} scheme, since ${\bf v}$ represents an exponentially weighted average of the previous values of ${\bf u}$. Now we repeat our previous approximation to this scheme. If, for example, the dynamics are solved using Euler's method, Eq.~(\ref{ev}) now becomes
\begin{align}
{\bf u}(t + \Delta t) &\approx {\bf g} - \alpha {\bf v}(t),\label{de1}\\
{\bf v}(t + \Delta t) &\approx {\bf v}(t) + \frac{\Delta t}{\tau}[{\bf u}(t) - {\bf v}(t)].\label{de2}
\end{align}
Again assuming constant ${\bf g}$, a linear stability analysis shows that when $\tau/\Delta t >1$ the fixed point ${\bf u} = {\bf v} ={\bf g}/(1+\alpha)$ is linearly stable as long as $\alpha < \tau/\Delta t$. While we don't expect this estimate to be exact, we expect that the range of values of $\alpha$ for which the control scheme is stable will be greatly expanded when $\tau/\Delta t$ is large. Interestingly, in contrast to typical control problems, the presence of delays increases the stability of the control scheme. In summary, the delayed control  algorithm for suppressing a disturbance ${\bf g}(t)$ is as follows: (i) Force the system with a known training forcing function ${\bf f}$, and train a reservoir computer so that its output ${\bf u}$  approximates ${\bf f}$ based on observations of the state variables $\hat {\bf x}$. (ii) Add a term $-\alpha {\bf v}$ to the disturbed system, where ${\bf v}$ satisfies Eq.~(\ref{control3}) with large $\tau$.

\begin{figure*}[t]
    \centering
    \includegraphics[width = 0.8\linewidth]{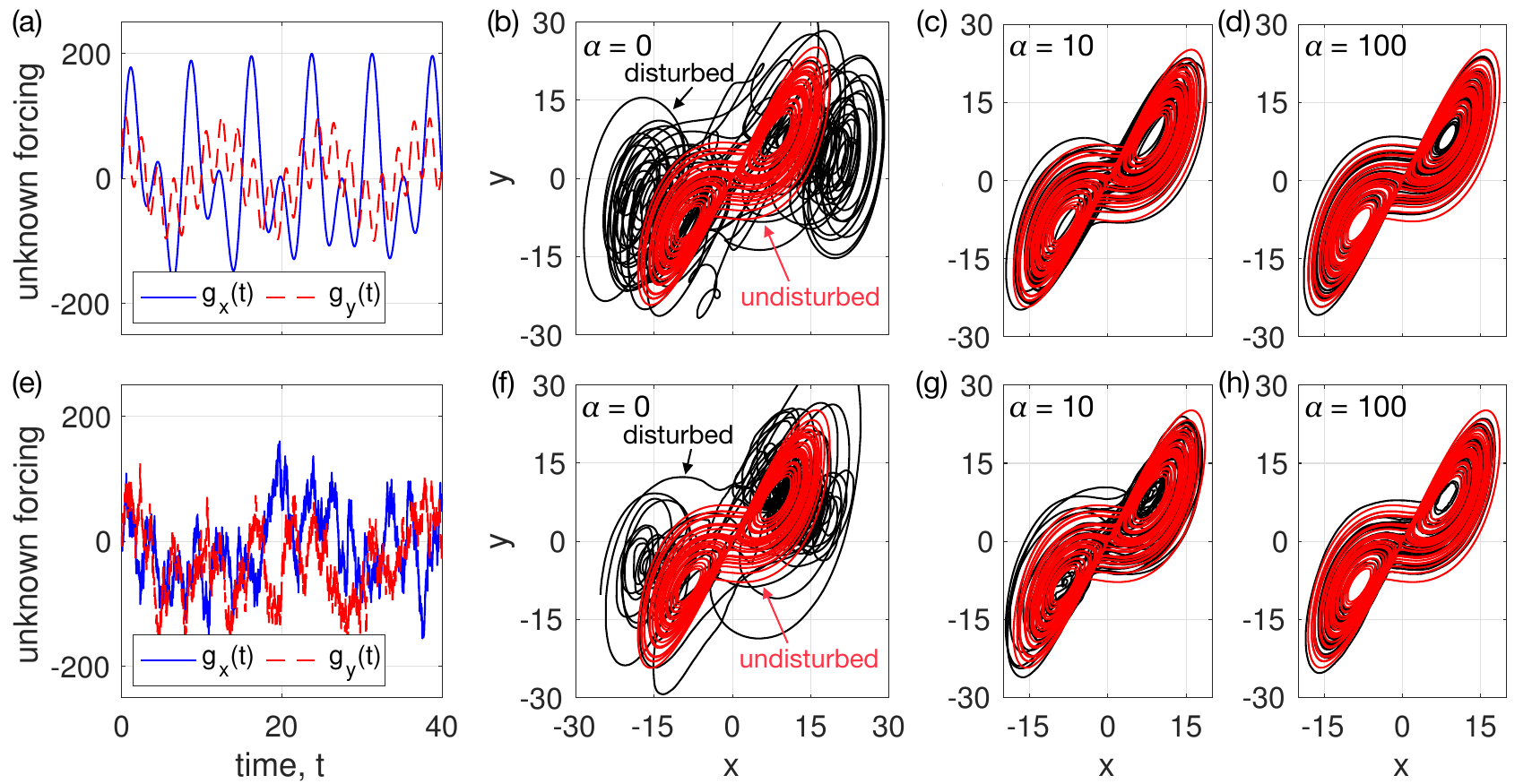}
    \caption{Suppressing unknown disturbances: Quasi-periodic and stochastic disturbances. For quasi-periodic (top) and stochastic (bottom) disturbances, the effectiveness of delayed control. In (a) and (e) the disturbances applied to the Lorenz system. For each case, (b)--(d) and (f)--(h) the disturbed (black) and undisturbed (red) attractors when using control gains $\alpha = 0$, $10$, and $100$.}
    \label{fig6}
\end{figure*}

\begin{figure*}[t]
    \centering
    \includegraphics[width = 0.78\linewidth]{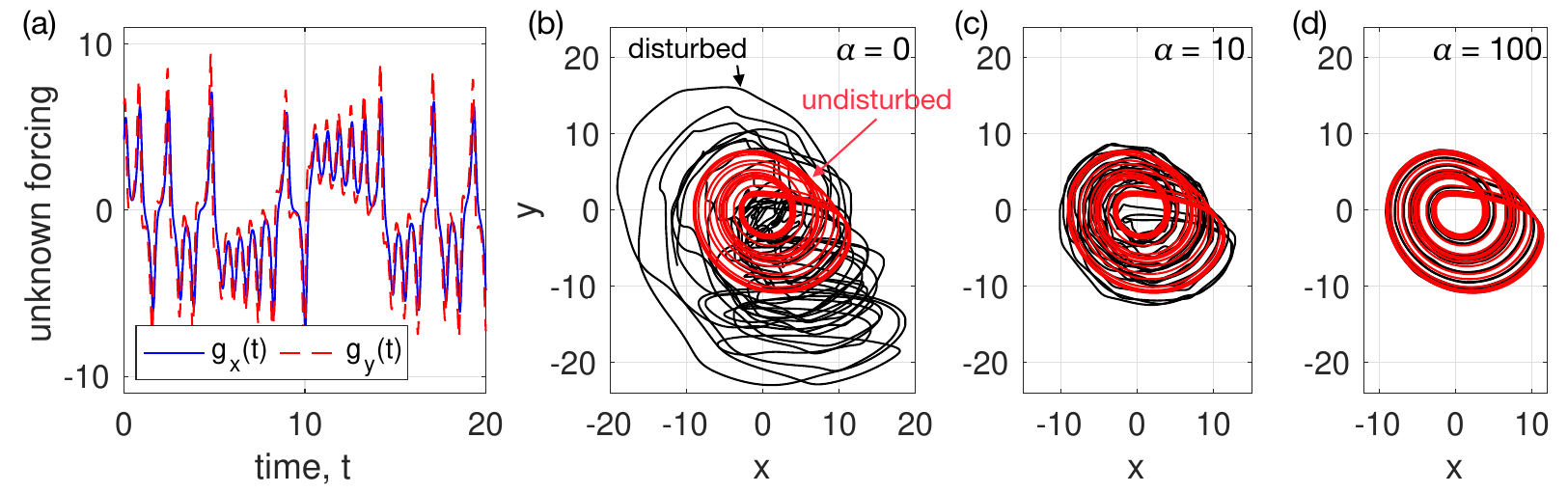}
    \caption{Suppressing unknown disturbances: R\"{o}ssler system. (a) The disturbances applied to the R\"{o}ssler system. (b)--(d) the disturbed (black) and undisturbed (red) attractors when using control gains $\alpha = 0$, $10$, and $100$.}
    \label{fig7}
\end{figure*}

\begin{figure*}[t]
    \centering
    \includegraphics[width = 0.80\linewidth]{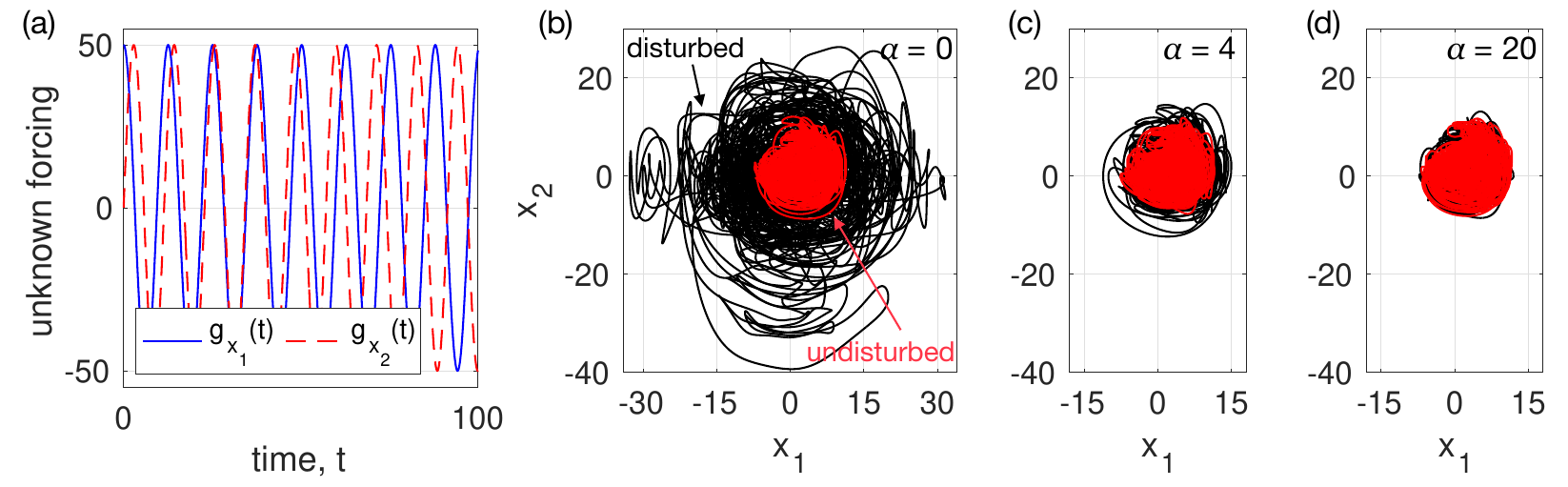}
    \caption{Suppressing unknown disturbances: Lorenz 96 system. (a) The disturbances applied to the Lorenz 96 system. (b)--(d) the disturbed (black) and undisturbed (red) attractors when using control gains $\alpha = 0$, $4$, and $20$.}
    \label{fig8}
\end{figure*}

\subsection*{Examples: Suppressing Deterministic Disturbances}\label{sec:02:04}

In order to demonstrate the suppression method discussed above, we return to our example of a Lorenz system forced by a R\"ossler system, except that the forcing applied to Eqs.~(\ref{l1})-(\ref{l3}), are greatly amplified, namely, $[g_x,g_y,g_z]^T = [24 x_R,24 y_R,0]^T$.
(In order to illustrate the power of our method, the disturbance terms are chosen to be much larger than in the previous example.) After training the reservoir with the sinusoidal forcing  $[f_x(t), f_y(t),f_z(t)]^T = [\cos(0.05t), \sin(0.05t),0]^T$, we run the control scheme in Eqs.~(\ref{control2})-(\ref{control3}). In Fig.~\ref{fig5}(a)--(c) we plot for control gains $\alpha=0$, $10$, and $100$ the disturbed Lorenz system in black curves as well as the undisturbed Lorenz system in red curves for comparison. Note that for $\alpha=0$ (i.e., no control) the disturbed system attractor bears little resemblance to the undisturbed system attractor, but as $\alpha$ is increased the control method begins to effectively mitigate the disturbances, with little effective difference for $\alpha = 100$. In order to quantify the effectiveness of the control method, we measure the distance between the disturbed and controlled attractor and the undisturbed attractor as follows. We solve Eqs.~(\ref{l1})-(\ref{l3}) with $g_x = g_y = g_z = 0$ for $T = 150$ time units using Euler's method with a timestep $\Delta t = 0.002$ after discarding a sizable transient and create a reference time series $\{{\bf x}_0(0),{\bf x}_0(\Delta t),{\bf x}_0(2\Delta t),\dots,{\bf x}_0(T/\Delta t)\}$ representing an approximation of the undisturbed attractor. Next, for a given value of $\alpha$, again after discarding a transient, we compute a time series for the disturbed and controlled system, $\{{\bf x}(0),{\bf x}(\Delta t),{\bf x}(2\Delta t),\dots,{\bf x}(T/\Delta t)\}$. Then we compute the average distance between the points on the controlled trajectory and the reference time series as
\begin{align}
d(\alpha) = \frac{1}{T /\Delta t} \sum_{i = 0}^{T/\Delta t} \min_j\| {\bf x}(i \Delta t) - {\bf x}_0(j \Delta t)\|.
\end{align}
In Fig.~\ref{fig5}(d) we plot the distance $d(\alpha)$ versus $\alpha$ for both the simple control scheme (\ref{control1}) (blue circles) and for the delayed control scheme (\ref{control2})-(\ref{control3}) (red crosses) for the deterministic disturbance. The simple control scheme reduces the error until it becomes unstable at approximately $\alpha \sim 2.5$. In contrast, the delayed control scheme reduces the error to very small levels for large values of $\alpha$, before it also becomes unstable at approximately $\alpha \sim 2500$. (For both methods, the values of $\alpha$ for which no data are shown resulted in numerical instability.) In practice, a suitable value of $\alpha$ could be chosen either by comparing the controlled attractor to the undisturbed one, if it is available, or by choosing $\alpha$ large enough that the controlled attractor doesn't change appreciably when increasing $\alpha$ further, as it is often done with the time-step of numerical ODE solvers.

We also present some additional results that demonstrate the generalizability of the suppression mechanism in the face of different types of disturbances. In particular, while we keep the undisturbed system defined by the Lorenz system, we consider first quasi-periodic disturbances that are composed from mismatched sinusoids, namely $[g_x,g_y,g_z]^T = [200\cos(2t/5)\sin(2\pi t/5),50\cos(t/2)+50\sin(\pi t),0]^T$, and, second, stochastic disturbances $[g_x,g_y,g_z]^T = [x_{S}(t),y_{S}(t),0]^T$ as defined in Eqs.~(\ref{eq:s1}) and (\ref{eq:s2}) but with a stronger stochasitc component, specifically $D=75$. We plot the results from these two cases under delayed control in Fig.~\ref{fig6} on the top and bottom, respectively. First, in panels (a) and (e) we plot the time series of the disturbances applied to the Lorenz system, depicting the disturbances to the $x$ and $y$ components in solid blue and dashed red, respectively. Next, in panels (b) and (f) we plot the disturbed attractor without control (i.e., using $\alpha = 0$) in black and also plotting the undisturbed attractor in red for comparison. Then in panels (c) and (g) we plot the disturbed and undisturbed attractors for $\alpha = 10$, then in panels (d) and (h) for $\alpha = 100$. As we increase the control gain $\alpha$ we see the disturbed attractor begins to resemble more so the undisturbed attractor. 

Lastly we consider control of disturbances to other systems, specifically the R\"{o}ssler system and the Lorenz 96 system with $N=8$ and $F=8$. In both cases we train the systems with sinusoidal forcing, $[f_x(t), f_y(t)]^T = [\cos(0.05t), \sin(0.05t)]^T$. Next, we disturb the $x$ and $y$ components of the R\"{o}ssler system with the $x$ and $y$ components of the Lorenz system, $[g_x,g_y]^T = [2x_{L(t)/5,2y_L(t)/5}]^T$ and we disturb the first two components of the Lorenz 96 system with sinusoids with offset frequencies, $[g_{x_1}(t), g_{x_2}(t)]^T = [50\cos(t/2),50\sin(11t/20)]^T$. In Figs.~\ref{fig7} and \ref{fig8} we plot the results for the R\"{o}ssler system and the Lorenz 96 system, respectively, plotting in panel (a) the disturbances applied to each, then in panels (b)--(d) the disturbed (black) and undisturbed (red) attractors as the control gain is increased: for the R\"{o}ssler system we use $\alpha = 0$, $10$, and $100$ and for the Lorenz 96 system we use $\alpha = 0$, $4$, and $20$. (Note that for the Lorenz 96 system we were able to suppress disturbances quite well with even smaller control gains, thus the smaller values of $\alpha$ used.) In both cases we see that as the control gain is increased the disturbed dynamics get closer to the undisturbed dynamics.

\section*{Discussion}\label{sec:03}

In summary, we have presented and demonstrated both numerically and experimentally a method that allows an unknown disturbance to an unknown dynamical system to be identified and suppressed in real-time, based only on previous observations of the system forced with a known forcing function. Our method is applicable, for example, to the problem of identifying node and line disturbances in networked dynamical systems such as power grids \cite{wang2019detection, delabays2021locating,delabays2022locating}, and more broadly to the various fields where disturbances need to be suppressed in real-time \cite{chen2015disturbance}. While our method does not require knowledge of the underlying dynamics of the system, it requires one to be able to force it with the addition of a known training forcing function, and subsequently with the term $-\alpha {\bf v}$. The consideration of nonlinear disturbances is left for another manuscript \cite{skardal2023detecting}. In addition, we assumed that all the variables of the system can be observed. In principle, one could use our method by training the reservoir using an observed function ${\bf H}({\bf x})$ of the state vector, but we have not explored this generalization. Another important research direction is to determine the class of appropriate training forcing functions, given a dynamical system and the anticipated characteristics of the disturbance. 

\section*{Methods}\label{sec:04}

\subsection*{Choice of the Bias Parameter, $\beta$}\label{sec:04:01}

\begin{figure}[t]
    \centering
    \includegraphics[width = 0.9\linewidth]{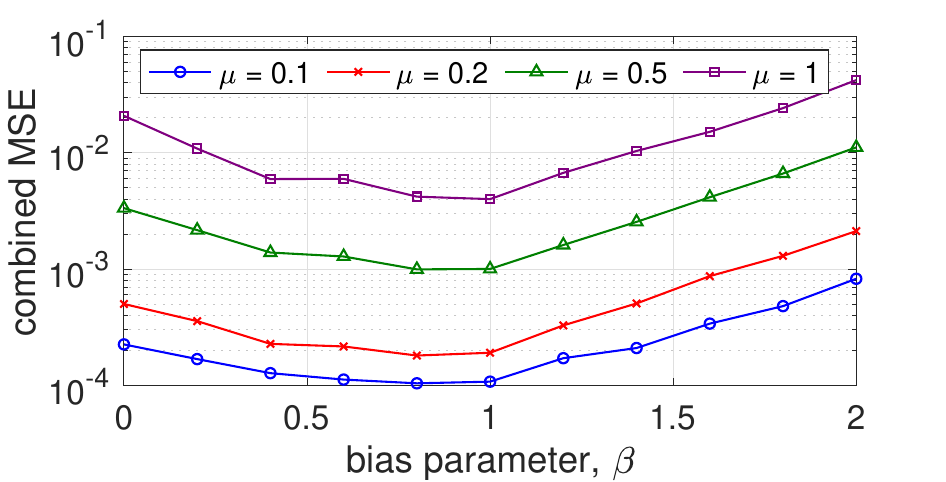}
    \caption{Effect of the bias parameter. As a function of the bias parameter $\beta$, the combined mean squared error (MSE) between the unknown disturbances, $g_x$ and $g_y$, and those recovered by the reservoir, $u_x$ and $u_y$. Unknown disturbances were presented at different magnitudes, scaled by $\mu$ as $[g_x,g_y,g_z]^T = [\mu x_R,\mu y_R,0]^T$.}
    \label{fig9}
\end{figure}

To explore the choice that the bias parameter $\beta$ [see Eq.~(\ref{eq:03})] has on the ability of the reservoir computer to recover unknown disturbances, we return to our first example of a Lorenz system with sinusoidal known forcing used as training and then disturbed by R\"{o}ssler dynamics. All other system and reservoir computer parameters are the same, training is set to $[f_x(t), f_y(t),f_z(t)]^T = [\cos(t/20), \sin(t/20),0]^T$, but we consider varying both the bias parameter $\beta$ and the magnitude of the R\"{o}ssler forcing, namely, we introduce a magnitude parameter $\mu$ that scales the unknown disturbance as $[g_x,g_y,g_z]^T = [\mu x_R,\mu y_R,0]^T$. (Note that in the main text we first used $\mu = 1/10$.) To examine the effect of the bias parameter we then train the reservoir with different $\beta$, then with that chosen value of $\beta$ try to extract the unknown disturbance at different levels of $\mu$. We evaluate the success of the reservoir in extracting the disturbances by calculating the sum of the mean squared error (MSE) in both the $x$ and $y$ components over the time window, namely, $\text{MSE}=(\int_0^{T}[g_x(t)-u_x(t)]^2 dt+\int_0^{T}[g_y(t)-u_y(t)]^2 dt)/T$. In Fig.~\ref{fig9} we plot the the combined MSE as a function of the bias parameter $\beta$ for a number of choices of the magnitude parameter, specifically $\mu=0.1$ (blue circles), $0.2$ (red crosses), $0.5$ (green triangles), and $1$ (purple squares). Results demonstrate that a bias parameter is optimal near $\beta =1$, which informs the choice made in this paper.

\subsection*{Experimental Implementation}\label{sec:04:02}

\begin{figure}[t]
    \centering
        \includegraphics[width = \linewidth]{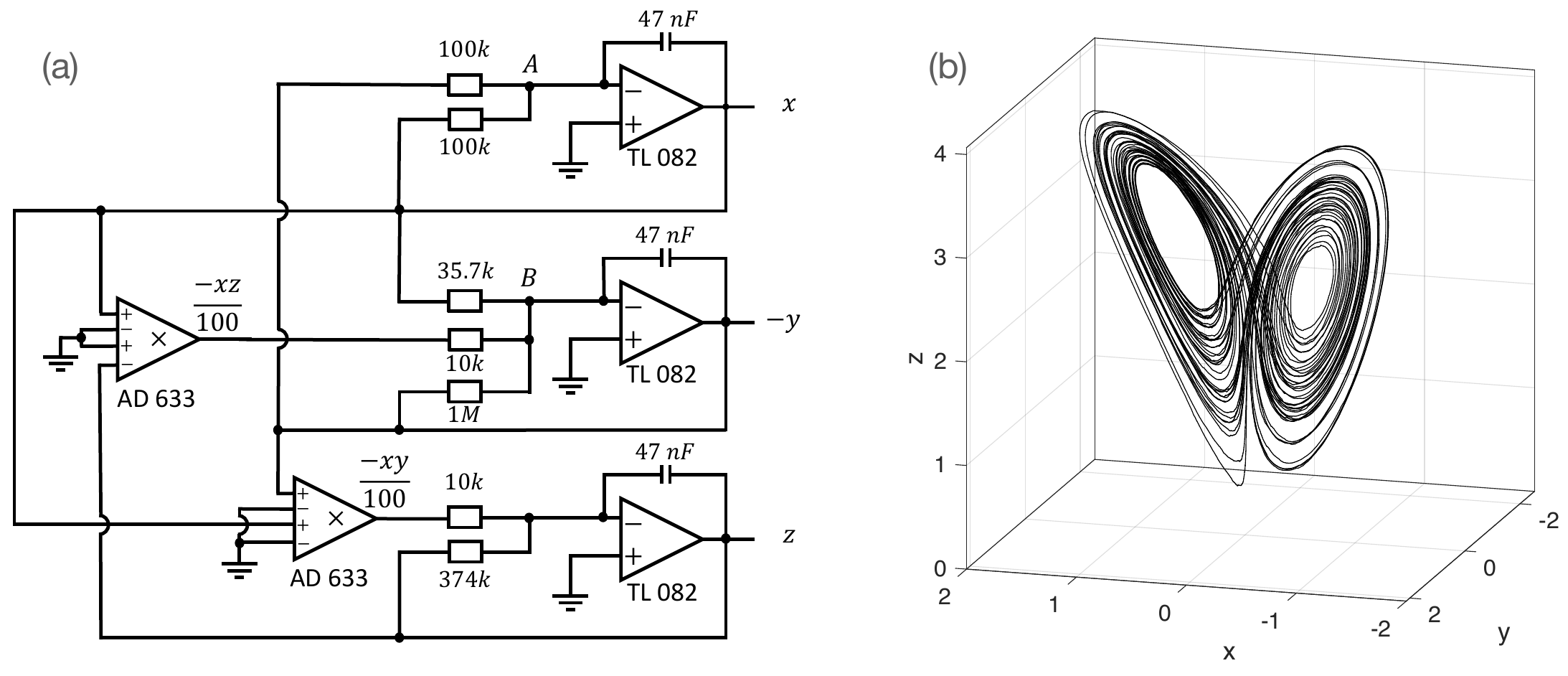}
     \caption{Circuit for Lorenz attractor shown on the left. Integrators are based on the TL082 operational amplifier while the multipliers are based on the AD633 chip. Circuit recreated from \cite{horowitz2020art}. Output of the circuit shown on the right, where units are in volts.}\label{fig:circuit}
\end{figure}

We constructed an analog electric circuit to replicate the Lorenz equations through three integrators and two multipliers, following the implementation described in Refs.~\cite{fitch2010analog, horowitz2020art} and shown schematically in Figure \ref{fig:circuit}(a). In this implementation, the variables $x$, $y$, and $z$ are the voltages shown in the diagram in Fig.~\ref{fig:circuit}(a) and correspond to the respective variables in the Lorenz system scaled down by a factor of $10$ (the equations the system models are modified accordingly for this scaling). 

The values used for the resistors are chosen to produce the appropriate coefficients in the Lorenz equations of $\sigma = 10$, $\beta = 8/3$, and $\rho =28$. The integrating capacitors of $47$ nF were chosen to provide oscillations on the order of 30 Hz. Resistors had a component tolerance of 1\%, while the capacitors have a 5\% tolerance. 
The multiplication was done with an AD633 analog multiplier, which has an error of $2\%$ of full scale, while the integrating circuit was based on an TL082 operational amplifier. The output of this circuit is shown in Fig.~\ref{fig:circuit}(b). The characteristic butterfly shape is readily apparent, with each output swinging around 4 volts peak to peak (Vpp). This analog circuit represents the ``undisturbed system'' described in the main text. While it is constructed to obey approximately the (scaled) Lorenz equations, the component tolerances make it, for practical purposes, an unknown system from which we can measure the state variables $x$, $y$, and $z$. Electronic noise and uncertainty from the analog multipliers adds an additional complication not present in our numerical simulations.

The external forcing is introduced into the circuit through the two points marked $A$ and $B$ in Figure \ref{fig:circuit}(a). A function generator produces two signals at magnitudes that were approximately 4 Vpp to closely match the magnitude of the variables of the unforced circuit. These forcing signals were each passed through a unity gain buffer and a 1 MOhm resistor before being added to the signal at the input of the x and y integrators at A and B, respectively. The value of 1 MOhm allows the forcing signal to be of a comparable amplitude to the x, y, and z signals, and adds this function unscaled into the first two integrators. 

\section*{Data Availability}
The datasets generated during and/or analyzed during the current study are available from the corresponding author on reasonable request.

\section*{Code Availability}
The code used during the current study is available from the corresponding author on reasonable request.

\acknowledgements
JGR acknowledges support from NSF Grant DMS-2205967. PSS acknowledges support from NSF grant MCB2126177.

\section*{Author Contributions}
JGR and PSS conceived the research. JGR, CPB, and PSS performed the research. JGR, CPB, and PSS wrote the manuscript.

\section*{Competing Interests}
The authors declare that they have no competing interests.




\bibliographystyle{unsrt}
\bibliography{refs} 


\end{document}